\documentstyle[12pt,epsf]{article}
\font\blackboard=msbm10 at 12pt
\font\blackboards=msbm7
\font\blackboardss=msbm5
\newfam\black
\textfont\black=\blackboard
\scriptfont\black=\blackboards
\scriptscriptfont\black=\blackboardss
\def\bb#1{{\fam\black\relax#1}}
\newlength{\dinwidth}
\newlength{\dinmargin}
\begin{document}
\newcommand{\ls}{\alpha'}
\newcommand{\cm}{\hspace{1cm}}
\newcommand{\z}{{\bb Z}}
\newcommand{\r}{{\bb R}}
\newcommand{\bc}{{\bb C}}
\newcommand{\be}{\begin{equation}}
\newcommand{\ee}{\end{equation}}
\newcommand{\ber}{\begin{eqnarray}}
\newcommand{\eer}{\end{eqnarray}}
\newcommand{\lp}{\left(}
\newcommand{\rp}{\right)}
\newcommand{\lk}{\left\{}
\newcommand{\rk}{\right\}}
\newcommand{\lc}{\left[}
\newcommand{\rc}{\right]}
\newcommand{\sT}{{\scriptscriptstyle T}}
\newcommand{\2}{\,\,2}
\def\a{\alpha}
\def\b{\beta}
\def\g{\gamma}
\newcommand{\se}{\section}
\newcommand{\tra}{\vec{p}_{\sT}}
\newcommand{\Z}{Z\left(\beta\right)}
\newcommand{\half}{\frac{1}{2}}
\newcommand{\pla}{\alpha'}
\thispagestyle{empty}

\begin{flushright}
\begin{tabular}{l}
FFUOV-99/03\\
{\tt hep-th/9903039}\\
\end{tabular}
\end{flushright}

\vspace*{2cm}

{\vbox{\centerline{{\Large{\bf ON THE MICROCANONICAL DESCRIPTION
}}}}}
\vskip30pt
{\vbox{\centerline{{\Large{\bf OF D-BRANE THERMODYNAMICS 
}}}}}
\vskip30pt

\centerline{Marco Laucelli Meana and Jes\'{u}s Puente
Pe\~{n}alba
\footnote{E-mail address:
    laucelli, jesus@string1.ciencias.uniovi.es}}

\vskip6pt
\centerline{{\it Dpto. de F\'{\i}sica, Universidad de Oviedo}}
\centerline{{\it Avda. Calvo Sotelo 18}}
\centerline{{\it E-33007 Oviedo, Asturias, Spain}}

\vskip .5in

\begin{center}
{\bf Abstract}
\end{center}  

   We study the microcanonical description of string gases in the presence
of D-branes. We obtain exact expressions for the single string density of
states and draw the regions in phase space where asymptotic approximations
are valid. We are able to describe the whole range of energies including the
SYM phase of the D-branes and we remark the importance of the infrared
cut-off used in the high energy approximations. With the complete expression
we can obtain the density of states of the multiple string gas and study its
thermal properties, showing that the Hagedorn temperature is maximum for
every system and there is never a phase transition whenever there is thermal
contact among the strings attached to different D-branes.


\newpage

\section{Introduction}

Finite temperature frameworks are in general very interesting laboratories
in which to study the  fundamental degrees of freedom of a theory. In the case
of string theory the interest increases because of the exponential growth
of the number of states with the mass. This behavior, known as the Hagedorn
spectrum \cite{hag}, is relevant because of the fact that it generates a
critical thermodynamics that in principle could correspond to a phase
transition \cite{carli,alv1,alv2,bowick}. The analysis of this possibility was 
firstly attempted from the canonical ensemble \cite{carli,alv1,alv2,bowick}, but it
seems that a more fundamental description of the degrees of the system is
needed.

The description that, in principle, would clarify the  open questions on the
Hagedorn thermodynamics is that of the microcanonical ensemble
\cite{deo,tan,us1,us2}. The density of states of the string gas was studied 
in the early works \cite{deo, tan} on the subject by  taking an ambiguous 
high energy approximation.  In fact the thermodynamics for the closed strings 
that was derived  there do not have any information about the low energy
behavior of the system. This prevents any trustable study on the way the
system reaches the Hagedorn temperature. 

On the other hand in \cite{deo, tan} there was a disregarding of the volume
dependence on the density of states that led to the surprising
conclusion that there is a difference between a system with an already open
universe and one in which we take to infinity the radii of the initial
torus \cite{tan}.

In \cite{us1,us2} another point of view was claimed. The density of states
of the single closed string valid for all energies was obtained. The multiple
string one was formally defined from the convolution theorem \cite{us1}, and
it was shown that for an already open universe the systems reaches the
Hagedorn temperature for a finite energy per string. At this point a phase
transition occurs in such a way that the temperature is kept constant and the
specific heat diverges. By this analysis it was possible to know how the
systems evolve from low energy to the high energy limit. 

The compactified case was studied in \cite{us2}. In this background the  system
has a unique phase with finite and positive specific case. The  Hagedorn 
temperature works as the limiting one. An interesting point was that in the
decompactification limit the system coincided with that of \cite{us1}.  

A common feature appearing in both cases is that at high energies  
equipartition breaks at a given energy per string. From this point the system 
is composed of a dominant highly  energetic long string in thermal  contact 
with a sea of a large number of low energetic strings that works as a thermal
bath. This relevant part of the system was not taken into account previously
in \cite{deo, tan} because it emerges from  the IR part of the single string
density of states. The existence of the string sea is relevant in order to 
prevent the negative specific heat phases. 

From an completely different perspective the same picture has been recently
obtained \cite{amati}.  The authors of \cite{amati}, considered the
possibility of a high energetic radiating string. They discovered that this
string would behave as a black body at a temperature $T_H$ and if one
included a thermal bath composed by low energetic strings the system would
reach equilibrium only if both temperatures, those of the black body and
those of the bath, coincide at the Hagedorn value. This strongly supports the idea
presented in \cite{us1,us2}, that a long energetic string would be in thermal 
equilibrium only in the presence of a thermal bath at $T_H$. 

To put it in short: working with approximate expressions without a neatly
defined range of validity is dangerous. It is difficult to discern in which
situations low energy effects can qualitatively change the conclusion of the
physical analysis. This is specially true in systems composed of many
objects, let alone those where a thermodynamical limit is to be taken and
even more near a phase transition, because local and small fluctuations
compared to the total energy can always make individual objects visit the
regions of phase space that have been thrown away. In our work we will
render an exact calculation of the single string density of states and a
comparison with the different approximation that come from it will be done. 

These considerations hold for perturbative string physics. It has
been sometimes argued that non-perturbative objects would have some role to
play in relation to the Hagedorn problem. There have been some works in
which this topic has been dealt with, either trying to include  D-branes
\cite{polchi} in the canonical ensemble \cite{green,vaz1,tse}, or
D-instantons \cite{vaz2} effects in the microcanonical ensemble. A laconic
'nothing happens' has been the usual conclusion. In the former case the
problem of approaching the Hagedorn problem from the canonical ensemble 
 is exactly the same that for closed strings. That is the canonical
description breaks and nothing else could be said since this point. In the latter
the main argument is that the energy scales at which the
non-perturbative effects become important are completely dominated by the
perturbative Hagedorn physics, so it seems natural to conclude that a
subsequent effect cannot regularize an already existing at low energies one.   
      
D-brane dynamics is governed by the Dirichlet open strings attached to it
\cite{polchi}, so in order to study D-brane thermodynamics we need to study
the finite temperature version of these open string theories
\cite{green,vaz1,tse}.
On the other hand  if we want to study the complete thermal behavior of type II
string theories we need to account for the closed string sector and for the
corresponding D-branes allowed for each theory. Finally if we want a
fundamental understanding of the Hagedorn 'phase' of the theory we need to
study it in the microcanonical ensemble. This idea has been recently
developed in \cite{barbo}. 

In this work  the microcanonical density of states has been computed for a
variety of brane configurations by using the methods of \cite{tan,deo}, and
high energy thermodynamics has been described. A related work was also made
in \cite{lee}.

In the present paper we approach the microcanonical description  of 
thermodynamics with D-branes as background  applying the methods of
\cite{us1,us2}. The plan of the work is the following. We begin the analysis
by computing the density of states from the mass spectrum. The next step 
consists of a computation of the single string density of states by the
method of the  inverse Laplace transform. We separate the analysis for
high energies, analogous to the analysis of \cite{barbo}, from those valid for 
the whole energy range. Using the previous results we go to the thermodynamical
 description of the system .

\section{The Density of States from the Mass Spectrum}

There are two ways to  compute the density of states of the single string
system.  The first one consists of the application of the inverse Laplace 
transform. The method begins with the computation of  the Helmholtz
free energy and then using it as the single string partition function in such 
 way that we obtain the density of state from the Bromwich integral
\be
\Omega^{(1)}(E)=\frac{1}{2\pi i}\int_{C}d\beta Z_1(\beta) e^{\beta E}
\ee
where $Z_1(\beta)=-V \beta F(\beta)$. The $n$-string term is characterized by the
partition function 
\be
Z_n(\beta)=\frac{1}{n!}Z_1(\beta)^n.  
\ee

The other way to calculate the microcanonical density of states of a system 
 is to count the degeneracy in the phase space geometrically. It is
specially useful in the case of compact target backgrounds.
In this section we shall implement this idea in our case of a gas of open
superstrings moving in the presence of D-branes. We start with the following
Hamiltonian, that corresponds to the mass formula of the theory which. We take
it as a dispersion relation

\be
\ls E^2=\vec{p}^2+\vec{\omega}^2+2 N
\ee
where $\vec{p}$ is the momentum, that is discrete in this case. The
dimension of this vector is the number of Neumann directions. $\vec{\omega}$   
is the winding number. It is always perpendicular to the momentum vector
because open strings cannot wind around a direction and have a momentum at
the same time. The dimension of this vector is, thus, the number of
Dirichlet directions. $N$ is the total oscillation number. It is the sum 
\be 
N=\sum n N_n.
\ee

  This means that there are several states with the same total oscillator
number. They are counted by the following coefficients:
\be
a_k=\frac{1}{2 \pi i} \oint \frac{d q}{q^{k+1}}
\lp \frac{
\vartheta_2(q)}{\eta^3(q)}\rp^{4-\frac{\nu}{2}}
\lp \frac{
\vartheta_3(q)}
{\vartheta_4(q)}\rp^{\frac{\nu}{2}}
\ee
where the integrand is the internal partition
function of the theory \footnote{
The dependence of the internal partition function of the theory on the
number of mixed, $ND$ and $DN$, direction comes from the different normal
ordering constant of these sectors. We are very grateful to J.L.F. Barb\'on
for focusing our attention on this point.}. 

   Let us begin to calculate the degeneracy associated to a value of the
energy fixing firstly $N$ to a constant. In that case:
\be
\vec{p}^2+\vec{\omega}^2=\ls E^2-2N=C
\ee

On the other hand
\be
\vec{p}^2+\vec{\omega}^2=\sum_{i\in NN} n_i^2 \frac{\ls}{R_i^2} + \sum_{i \in
DD}
\frac{R_i^2}{\ls}
\ee
where $NN$ marks the directions with Neumann boundary conditions and $DD$ the
Dirichlet ones, there are also mixed directions, $DN+ND=\nu$, that do not contribute to
the kinematical energy but through their oscillator masses only. This
contribution is accounted by the coefficients $a_N$ . To make things more
 homogeneous, we can use the variables
\ber
r_i=\frac{R_i}{\sqrt{\ls}}  \hspace{2cm} \mbox{if $i$ is Neumann} \nonumber
\\
r_i=\frac{\sqrt{\ls}}{R_i}  \hspace{2cm} \mbox{if $i$ is Dirichlet}
\eer
so that
\be
\vec{p}^2+\vec{\omega}^2=\sum_i \frac{n_i^2}{r_i^2}=C
\label{ellip}
\ee
this is the equation of an ellipsoid in the discrete phase space. We are
interested in counting the number of states for which that relation
approximately holds, that is, letting the energy be in $[E-\Delta E,E+\Delta
E]$. If the energies are very low, it is important to remind that $\Delta E$
has a lower bound which is the maximum gap between two energy levels. The
number of states is proportional to the volume of the ellipsoidal shell
defined by equation (\ref{ellip}). The total volume is
\be
V(C)=\Omega_{\tilde{D}} \lp \prod_i r_i \rp C^{\tilde{D}/2}.
\ee
The volume of a thin shell is
\ber
V_{\mbox{\tiny shell}}=\Omega_{\tilde{D}} \frac{\tilde{D}}{2} \lp \prod_i r_i \rp
C^{\frac{\tilde{D}}{2}-1}
\Delta C= \nonumber \\
=\Omega_{\tilde{D}} \tilde{D}  \lp \prod_i r_i \rp \lp \ls E^2-2k
\rp^{\frac{\tilde{D}}{2}-1}
\label{shell}
\ls E \Delta E
\eer
where $\Omega_{\tilde{D}}$ is the angular part. Its dependence on the dimension is
\be
\Omega_{\tilde{D}}=\frac{2 \pi^{\tilde{D}/2}}{\tilde{D} \Gamma(\tilde{D}/2)}
\ee
  In our case $\tilde{D}=DD+NN\leq 9-\nu$, is the number of testable
directions.

  To get the total degeneracy it is necessary to multiply by the one
associated to $N$ ($a_N$) and by the size of the super-multiplet: $2^4$.
After the sum over all possible values of $N$ is performed, the result is
\be
\Omega_{\mbox{\tiny cl}}(V,E)=2^4 \frac{2
\pi^{\tilde{D}/2}}{\Gamma(\tilde{D}/2)} (2
\pi)^{DF-NN}
\frac{V_{NN}}{V_{DD}} \sum_{N=0}^\infty  a _N \ls E
\lp \ls E^2-2N \rp^{\frac{\tilde{D}}{2}-1} {\cal H}\lp \ls
E^2-2N \rp
\ee

   This is not valid for $\tilde{D}=0$ because the derivative in
(\ref{shell}) is not correct. In that case there is no continuous
contribution from either the momenta or the windings and do the density of
states is just a sum of delta functions centered on the mass levels. We
shall talk further in section $5$.

    For a value of $E$, the sum over $N$ is finite because the oscillator
number is limited by the available energy. This is expressed by the
Heaviside function. The calculation will be complete when we take into
account the quantum statistics. Before doing that, let us add some terms
that are important if the proportion between some radii is very different
from one. In that case, there is a range of energies where the first
non-zero modes of some directions lie far beyond the ellipsoid in the phase
space and the geometric count is no longer valid because the single zero
mode is not counted as one but as a small fractional number of states. To
simplify, we shall consider all the Neumann radii equal and the same with
the Dirichlet ones. The only parameters are then $R_{DD}$ and $R_{NN}$. This
can be generalized to consider all the possibilities but there are too many
and all the interesting physical consequences can be extracted studying this
simpler case. The density of states that this yields is

\ber
\Omega_{\mbox{cl}}(V,E)=2^4 \frac{2 \pi^{\tilde{D}/2}}{\Gamma(\tilde{D}/2)} (2 \pi)^{DD-NN}
\frac{V_{NN}}{V_{DD}} \sum_{N=0}^\infty  a _N \ls E \times \nonumber \\
\times \lp \ls E^2-2N \rp^{\frac{\tilde{D}}{2}-1} {\cal H}\lp \ls
E^2-2N \rp  {\cal H}\lp E- V_C^{-1/C} \rp  + \nonumber \\
+2^4 \frac{2 \pi^{C/2}}{\Gamma(C/2)}
\frac{(2 \pi)^C}{V_C} \sum_{N=0}^\infty  a _N \ls E
\lp \ls E^2-2N \rp^{\frac{C}{2}-1} {\cal H}\lp \ls
E^2-2N \rp  {\cal H}\lp V_C^{-1/C}-E  \rp
\eer
   where
\ber
\mbox{If} \cm R_{NN}\gg 1/R_{DD} \cm & C & \cm \mbox{stands for $NN$ and} \cm
V_C=V_{NN}
\nonumber \\
\mbox{If} \cm R_{NN}<<1/R_{DD} \cm & C & \cm \mbox{stands for $DD$ and} \cm
V_C=1/V_{DD}
\eer
   Now it is easy to add quantum statistics following this recipe
\ber
-\beta F(\beta,V)_{B,F}=\pm\sum_{\{ n\} } log\lp 1 \mp e^{\beta E(\{ n\}
)}\rp=
\nonumber \\
=\mp \sum_{\{ n\} } \sum_{r=1}^\infty \frac{(\pm 1)^r}{r} e^{\beta E(\{ n\}
)}
\eer

 This leads to the final result

\ber
\Omega(V,E)=2^4 \frac{2 \pi^{\tilde{D}/2}}{\Gamma(\tilde{D}/2)} (2 \pi)^{DD-NN}
\frac{V_{NN}}{V_{DD}} \sum_{N=0}^\infty \sum_{r\ge 1,\mbox{odd}}^\infty
\frac{1}{r^{\tilde{D}+1}} a _N \ls E \times \nonumber \\
\times \lp \ls E^2-2N r^2 \rp^{\frac{\tilde{D}}{2}-1} {\cal H}\lp \ls
E^2-2N r^2 \rp  + \nonumber \\
+2^4 \frac{2 \pi^{C/2}}{\Gamma(C/2)}
\frac{(2 \pi)^C}{V_C} \sum_{N=0}^\infty  a _N \ls E
\lp \ls E^2-2N \rp^{\frac{C}{2}-1} {\cal H}\lp \ls
E^2-2N \rp  {\cal H}\lp V_C^{-1/C}-E  \rp.
\eer
This result is completely analogous to that of \cite{us2}. We will see in
next sections how we can recover the same density of states for the single
string system from the study of the free energy of the theory in the
canonical ensemble.

\section{High Energy Density of States}

In this section we are going to compute the Helmholtz free energy of the Dirichlet open
string theory for a variety of D-brane backgrounds from the canonical ensemble, and we 
will finally study the  different limits of the density of states derived
from it. We will do it for the asymptotic high energy range. 
The approximation is based on the idea
that the existence of the Hagedorn critical temperature is an UV phenomenon so 
it should be enough to look at the very excited string states to obtain the 
relevant information, this is the perspective assumed in \cite{tan,deo}, and 
more recently, for the D-brane backgrounds, in \cite{barbo}. The other 
option is to try to compute the complete density of states without
projecting out any string states, and will be detailed in the section 5.

We will begin by  noticing  that the degrees  of freedom lost in this high-energy
limit, are, in fact, projected out in the canonical description by a  IR 
cut-off that cannot be forgotten. If we introduce it explicitly, we would easily 
see how the massless open strings are neglected. This projection prevents the possibility of 
accounting for the low energetic sea. 

Let us firstly fix our background. Suppose we have a configuration with a
Dp-brane and a Dq-brane. We can define four different types of directions,
the Dirichlet-Dirichlet(DD),Neumann-Neumann (NN), ND and DN. The last two
do not contribute to the mass of the strings, apart from the oscillation
terms (That means that we can neither have momentum nor winding in these
directions).
We will include in this classification of the directions the spatial ones
only. They are related by 
\be
DD+NN+DN+ND=9.
\ee
The Euclidean time is supposed to be compactified on a circle of length
$\beta$. Another interesting parameter we will deal with is the sum of the
mixed directions, $\nu=ND+DN$. The interest of it comes from the fact that the
when $\nu =0,4,8 $ the string gas is evolving in  a completely supersymmetric background. We
will focus our study on this type of brane configuration, however some
comments on more  general ones will be made.

In brief, the string's quantum numbers are: the momentum mode in the NN
directions; the winding mode in the DD ones and the Matsubara frequency in
the Euclidean time, the DN and ND direction do not contribute to the mass of
the strings.   

We can now compute the Helmholtz free energy as the vacuum energy of the theory
described above. It formally reads 
\be
-\beta F(\beta)=- Tr_{H} (-1)^F\int_0^{\infty}\frac{dt}{2t}e^{-2 \pi \pla t M^2}
\ee
where $M$ is the mass of the string and is given by
\be
M^2=\sum_{i=1}^{NN} \frac{n^2_{i}}{R_i^2}+\sum_{i=1}^{DD} w^2_i R_i^2+\frac{4\pi^2
m^2}{\beta^2}+Osc
\ee
where the last contribution comes from the oscillator modes of the open
string. We can compute the trace over the Hilbert space  obtaining
\be
-\beta F(\beta)=-\int_0^{\infty}\frac{dt}{2t} \prod_{i}^{NN} \theta_3 \lp 0,
\frac{2 i t \pla}{R_i^2} \rp  \prod_{i}^{DD} \theta_3 \lp 0, 
2 i t \pla R_i^2 \rp \theta_4 \lp \frac{8 \pi^2  i t \pla}{\beta^2}
\rp f(t) 
\label{complete}
\ee
where the internal partition function $f(t)$ is given by
\be
f(t)= \lp \frac{
\vartheta_2(0,it)}{\eta^3(it)}\rp^{4-\frac{\nu}{2}}
\lp \frac{
\vartheta_3(0,it)}
{\vartheta_4(0,it)}\rp^{\frac{\nu}{2}}.
\label{internal}
\ee

The high energy behavior of the open string gas may be studied in the
canonical ensemble and that is the easiest way of visualizing the Hagedorn
critical behavior. Let us begin our study by looking at this region. When the
energy is large the string modes that  dominate the dynamics will be the
massive ones. These are enclosed in the UV region of the previous
expression, that is accounted by the large $t$ limit. We can introduce an IR
cut-off $\Lambda^2$ in (\ref{complete}), and in this way we will forget about
low energetic strings, we have

\be
-\beta F(\beta)=-\int_0^{\Lambda^2}\frac{dt}{2t} \prod_{i}^{NN} \theta_3 \lp 0, 
\frac{2 i t \pla}{R_i^2} \rp  \prod_{i}^{DD} \theta_3 \lp 0, 2 i t \pla R_i^2 
\rp \theta_4 \lp \frac{8 \pi^2  i t \pla}{\beta^2} \rp f(t)
\label{cut-off}
\ee

In order to study the UV limit ($t\rightarrow 0$) it is convenient to take advantage of modular
properties of the theta functions in the free energy going to the
closed-string channel by making $t=2 \pi^{2}s^{-1}$, obtaining
\ber
-\beta
F(\beta)\sim -\frac{V^{NN}}{V^{DD}}\frac{\beta}{4\pi} 
\int_{\frac{2 \pi^{2}}{\Lambda^2}}^{\infty} ds
\prod_{j}^{NN} \theta_3\lp 0,\frac{i R_j^2 s}{4 \pi^2 \pla} \rp 
\prod_{j}^{DD} \theta_3\lp 0,\frac{i s}{4 \pi^2 R_j^2 \pla} \rp \nonumber\\
\times
 \theta_4 \lp 0,\frac{i \beta^2 s}{16 \pi^3 \pla} \rp
\lp \frac{
\vartheta_4(0,\frac{is}{2\pi^2})}{\eta^3(\frac{is}{2\pi^2})}\rp^{4-\frac{\nu}{2}}
\lp \frac{
\vartheta_3(0,\frac{is}{2\pi^2})}
{\vartheta_2(0,\frac{is}{2\pi^2})}\rp^{\frac{\nu}{2}}
\label{hag1}
\eer

We can now go to the large $s$ limit by simply taking into account that 
\be
\lim_{s\rightarrow\infty}
\lp \frac{
\vartheta_4(0,\frac{is}{2\pi^2})}{\eta^3(\frac{is}{2\pi^2})}\rp^{4-\frac{\nu}{2}}
\lp \frac{
\vartheta_3(0,\frac{is}{2\pi^2})}
{\vartheta_2(0,\frac{is}{2\pi^2})}\rp^{\frac{\nu}{2}}=e^{s/2\pi},
\label{tachyon}
\ee
we finally can express the UV limit of the open string free energy, with the
correct multiplicative prefactors, as
\be
-\beta
F(\beta)=-\frac{V^{NN}}{V^{DD}}(2\pi)^{DD-NN} (2 \pla)^{\frac{\nu-10}{2}}
\frac{\beta}{4\pi} (2\pi^2)^{-1}
\sum_{\{M,n\}}\int_{\frac{2 \pi^{2}}{\Lambda^2}}^{\infty} ds\, e^{-s \Delta
(M,n,\beta)}
\ee
where we have defined
\be
\Delta(M,n,\beta) = M^2+\frac{n^2 \beta^2}{16 \pi^2 \pla}-\frac{1}{2\pi},
\ee
in that way the winding and momentum modes are included into the mass $M$.
In what follows we will collect the numerical factors into a multiplicative
constant $C$.   
When $\Delta(M,n,\beta) =0$ we arrive at a critical temperature analogous to
the Hagedorn one. In fact we can see that at those temperatures a string
state becomes massless. When the temperatures are passed through this states
get a negative squared mass, that is, it  becomes tachyonic. We can solve the
equation for the critical temperatures obtaining 
\be
\beta_c^2=\frac{8 \pi \pla}{n^2}-\frac{16 \pi^2 \pla M^2}{n^2} \,\,\,\,\,
\hbox{with} \,\,\,\, 0<n \in \z.
\label{critical}
\ee

With $n=1$ and $M=0$ we have the lowest critical temperature that
corresponds to the Hagedorn one. The following critical points are obtained
for bigger temperatures, in fact we can in principle  have negative
unphysical solutions
for $\beta_c$. This possibility is avoided by assuming that the
compactification radii are bigger than the selfdual ones.   

At this moment we can write the Helmholtz free energy as a sum over
functions which are analytical in the whole complex-$\beta$ plane except in
the neighborhood of $\beta_c$. The sum reads

\be
-\beta
F(\beta)=- C \frac{V^{NN}}{V^{DD}}\frac{\beta}{4\pi}
\sum_{\{\beta_c,n\}}\int_{\frac{2 \pi^{2}}{\Lambda^2}}^{\infty} ds
\,e^{-\frac{s n^2}{2\pi}\lp\frac{ \beta^2-\beta_c^2 }{ \beta_H^2}\rp}.
\label{freec}
\ee

Finally we can integrate the previous expression in order to obtain a closed
expression for the free energy in the high energy limit. Using the integral
representation of the  incomplete Gamma function we have
\be
-\beta
F(\beta)=
-C\frac{V^{NN}}{V^{DD}}\frac{\beta}{4\pi}                
\sum_{\{\beta_c,n\}}
\lc
\frac{n^2}{2\pi} 
\lp
\frac{\beta^2-\beta_c^2}{\beta^2_H} 
\rp 
\rc^{-1}
\Gamma 
\lp 
1,\frac{\pi n^2}{\Lambda^2}
\lp
\frac{\beta^2-\beta_c^2}{\beta^2_H}
\rp
\rp.
\label{asymptotic}
\ee

The problem of obtaining the single string density of states from the free
energy below is reduced to an inverse Laplace transformation. It is known
that the result of this kind of computation is closely related with the
analytical properties of the function to be transformed. We can look at
(\ref{asymptotic}) as a sum of independent terms seeing that each one is
characterized by its behavior in the neighborhood of its thermal critical
point $\beta_c$. We see that the analytical properties are enclosed in the
simple pole and in the more complicate properties of the incomplete Gamma
function. The computation of the density of states here would consist of a
straightforward adaptation of those in \cite{us1}.

Before showing the results of the computation let us comment some words about the cut-off
$\Lambda^2$. We can try to extend the validity of the approximation taking
the value of $\Lambda^2$ to infinity. It will give us the complete version
of the Gamma function. The analytical
properties of the Free Energy are enclosed inside the single pole appearing
at each critical temperature. If we complete the IR part of the Free Energy
by simply taking $\Lambda^2$ to infinity we will obtain an IR regular series
for the density of states of the single string. The thermodynamics derived
from it would have a positive and finite specific heat phase only, in which the
asymptotic maximum temperature is the Hagedorn one. We think that this
corresponds to a misconception. The IR behavior that one could obtain from the previous
procedure is at least not complete. It is due to the fact that if we took a
large $\Lambda^2$ limit we could not truncate the string theory partition
function as we did in (\ref{tachyon}), but we should have included all the
string modes.

We think that the correct way of using
the previous free energies consist of keeping  $\Lambda^2$ finite, and
making inverse Laplace transformation of the Laurent series around the
critical points. Doing it we will see that we obtain an
asymptotic series that does not work for the IR regime. 

Following the previous  prescriptions we can make the same computation of the high 
energy single string density of states as in 
\cite{us1}, it  gives
\be
\Omega^{(1)}(E)=\frac{V^{NN}}{V^{DD}}\sum_{\{\beta_c,n\}}\sum_{p=0}^1\sum_{m=0}^{\infty}
A(\beta_c,n,m,p)\frac{(-1)^{m+p}e^{\beta_c E}}{E^{m+p}} {\cal H}\lp
E-\frac{2 \pi \beta_c n^2}{\beta_H^2 \Lambda^2}\rp
\label{omegaHigh}
\ee
where $A(\beta_c,n,m,p)$ is a constant independent of $E$ but dependent,
through $\beta_c$, of the $R$ parameters, and $ {\cal H}(x)$ is the Heaviside 
function. The sum in $m$ comes from a Taylor expansion of the incomplete Gamma 
function.

The behavior is in general  governed by a leading term that corresponds to 
the Hagedorn singularity. We will come back to the details later but let us 
sketch the general dynamics. We find a positive, but infinite, specific heat 
phase. The leading  term
does not give any negative  powers of the energy taking us to theory with a
constant temperature  and infinite specific heat behavior. The subleading
terms give corrections for small energies. The  string gas 
takes a finite and positive specific heat phase.

We think that these corrections are meaningless because the density of
states is only valid for a very high energy, and in order to know the low
energetic phase we need an IR-complete density of states.  We will come back
on these features later when studying the thermodynamical behavior of the
gas.

It seems  interesting to study the high energy density of states when we
open some of the DD directions. This case physically means that there would
be a reduced number of testable directions for the single string and has
shown its interest in previous studies \cite{deo,barbo,riotto}. The result
is analogous to  (\ref{omegaHigh}). It is implemented  by taking some of 
 the $DD$, (let it be a number $0\leq a \leq DD$ of them), infinite. We will
also suppose that the momenta on the $NN$ directions are continuous
so
\be
R^{DD}_{1},...,R^{DD}_{a},R^{NN}_{1},...,R^{NN}_{NN}\rightarrow \infty.
\ee
Including it into (\ref{hag1}), and then also in(\ref{critical}), we arrive at 
the following UV canonical Free Energy
\be
-\beta
F(\beta)=
-C\frac{V^{NN}}{V^{DD-a}}\frac{\beta}{4\pi}                
\sum_{\{\beta_c,n\}}
\lc
\frac{n^2}{2\pi} 
\frac{(\beta^2-\beta_c^2)}{\beta^2_H} 
\rc^{\frac{(9-\tilde{D}-\nu)}{2}-1}
\Gamma 
\lp 
1+\frac{\tilde{D}+\nu-9}{2}
,\frac{\pi n^2}{\Lambda^2}
\frac{(\beta^2-\beta_c^2)}{\beta^2_H}
\rp.
\label{asymptoticEFF}
\ee
where $\tilde{D}$ is defined as a effective number of testable dimensions.
It is given by

\be
\tilde{D}=(DD-a)+NN\leq 9- \nu.
\ee
The meaning of this number of dimensions and the existence of 'testable'
energy scales will be discussed later.

In this case the ananlytical properties are more complicate. In general we
can see that we have a branch point located at each critical temperature,
concretely it will happens for an odd number of untestable $DD$ directions,
while for even ones we would have high order pole singularities. The IR
continuation, by taking $\Lambda^2$ to infinity, is not well defined here.
The density of states that results from inverse Laplace transform
(\ref{asymptoticEFF}) is 
\be
\Omega^{(1)}(E)=\frac{V^{NN}}{V^{DD-a}}\sum_{\{\beta_c,n\}}\sum_{p=0}^1\sum_{m=0}^{\infty}   
A(\beta_c,n,m,p)\frac{(-1)^{m+p}e^{\beta_c E}}{E^{(9-\nu-\tilde{D})/2+m+p}} {\cal H}\lp
E-\frac{2 \pi \beta_c n^2}{\beta_H^2 \Lambda^2}\rp.
\label{omegaD}
\ee
We will see in the next section that these single string systems have a negative specific
heat phase only.

 The marginal case corresponds to opening all the $DD$
direction obtaining the density of states of a pure $NN$ theory 
\be
\Omega^{(1)}(E)=V^{NN}\sum_{\{\beta_c,n\}}\sum_{p=0}^1\sum_{m=0}^{\infty}
A(\beta_c,n,m,p)\frac{(-1)^{m+p}e^{\beta_c E}}{E^{(9-\nu-NN)/2+m+p}} {\cal H}\lp
E-\frac{2 \pi \beta_c n^2}{\beta_H^2 \Lambda^2}\rp.
\label{omegaHighopen}
\ee
This result describes a purely Neumann open string theory in $NN$
target dimensions. The energy dependence of the thermodynamics in the
microcanonical ensemble exactly matches with a certain brane configuration
described by (\ref{omegaD}), but without any volume dependence of the $DD$
side.  

In (\ref{omegaHigh}) and following, we can see that the density of states for the single
string system obtained with this procedure is useless in the IR regime
because of the use of the IR cut-off. This property prevents the use of this
magnitude for the computation of the multiple string system by the
convolution procedure \cite{us1}.

\section{High Energies Thermodynamics}

In this part of our work we will go on the thermal properties of the single
string system by using the density of states in (\ref{omegaD}). We will
begin with the analysis of the ranges of validity and we finally try to
extract from it some results on the multiple string thermodynamics. 

In this section we will be interested in comparing our results with those of
\cite{barbo} and we will find where their asymptotic approach breaks down
making  the convolution approach of \cite{us1,us2} indispensable.  

Let us begin by simply computing the microcanonical temperature. At first
order it reads  as 
\be
\beta=\frac{\partial}{\partial E}
\log \Omega^{(1)}_{\tilde{D}}(E)\sim \frac{\partial}{\partial E}\lp \beta_H
E-\lp \frac{9-\tilde{D}-\nu}{2} \rp\log E \rp {\cal H}\lp E-\frac{2 \pi}{\beta_H
\Lambda^2}\rp .
\label{microtemp}
\ee
Where we have  assumed that the dominance of the term corresponding to the
Hagedorn singularity. The properties that we can derive from it  hold only 
at high energies. We will detail the analysis of the validity range in
detail later. 

The temperature that results from (\ref{microtemp})   are drawn in the 
Fig.(1), where we have plotted the temperature versus the 
energy of the string. We can see that  the systems
have  negative specific heat phase in all cases except for $\tilde{D}+\nu=9$ in
which the entropy is proportional, at leading order, to the energy. In this
case the system has an infinite specific heat phase only, that corresponds to a
constant temperature. This is the only case in which the Hagedorn
temperature plays the role of  limiting temperature for the single string.
The extrapolation for the multiple string system must be done
carefully\footnote{In order to connect with the notation of \cite{barbo}
remember that 
\be
\gamma=\frac{9-\tilde{D}-\nu}{2}-1.
\ee}.
\begin{figure}
%
%
\let\picnaturalsize=N
\def\picsize{2in}
\def\picfilename{temp.ps}
\ifx\nopictures Y\else{\ifx\epsfloaded Y\else\input epsf \fi
\let\epsfloaded=Y    
\centerline{\ifx\picnaturalsize N\epsfxsize \picsize\fi
\epsfbox{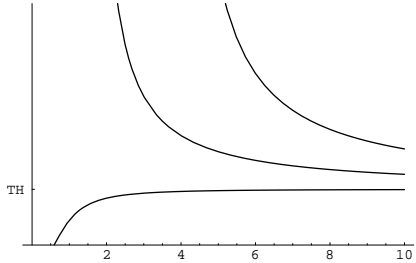}}}\fi
\caption{{\small Temperature versus energy of the various  single string system. 
We see that we have positive specific heat for the $\tilde{D}+\nu=9$ case
only, that corresponds to a system that tests the whole target space.}}
\label{temp1}
\end{figure}
We have introduced the effective dimension $\tilde{D}$ that correspond to
the number of dimensions the system can probe. In fact we have the $NN$
dimensions, that will be supposed to give a continuous momentum spectrum,
and $DD-a$ Dirichlet dimensions the string can wrap around. To eliminate
possible winding excitations around $a$ Dirichlet  directions we have already taken the
corresponding radii to infinity. However we could wonder that the system has
not enough energy to reach a winding state even when the radii are finite.
Physically this situation can be clearly explained  for the single string system but it comes
to us more subtle for the multiple string gas. In fact in order to
prevent the winding excitations one has to take the energy lower than
$R_a^2$. If we impose it for the multiple string gas it would reduce the
system to evolve in a very low energetic phase.

Let us see it explicitly. If we introduce the parameters $\rho_N$, that
corresponds to the energy per string, and the number of strings per unit of
$NN$ volume $n$, we conclude that
\be
\rho_N< \frac{R_a}{n V^{NN}}=\rho_{\omega}.  
\label{nowindingenergy}
\ee
Then if we want to take a thermodynamic limit ($V^{NN}\rightarrow \infty$,
and all the densities finite), we need to take the $a$ radii to infinity
too.

In summary, if we want to make a dimension untestable for the multiple
string system we should to take its radius to infinity, if we did not do it 
we might be able to talk about reduced effective dimension at very low energy 
densities $\rho_N$.  In order to make our work complete we will make the analysis for 
string gases with large, but finite, $R_a$ radius specially focusing our 
attention on the its limitation. 

Firstly we need to find the energy ranges in which the term
corresponding to the Hagedorn critical point in (\ref{omegaD}) dominates the
thermal properties. It will correspond to those energies at which the
subleading terms are negligible, that is
\be
e^{-(\beta_H-\beta_c)E}\sim {\cal O}(0)\Rightarrow (\beta_H-\beta_c)E \gg 1.
\label{hdr}
\ee
On the other hand we need to remember that the Hagedorn  term  in 
(\ref{omegaD}) is null for
\be
E<\frac{2\pi}{\beta_H \Lambda^2}=E_c
\ee 
where this condition comes from the Heaviside function. The combination of
the previous expression gives the range of energies at which the system
has a simply Hagedorn behavior. In order to specify it we need to know how
is the dependence of (\ref{hdr}) on $\beta_c$. 

We know that the critical points are given by (\ref{critical}). Because of 
the assumption that momenta should be continuous we only deal with critical 
points coming from winding modes and high-mass levels of the string. For the latter we
conclude that the second critical point $\beta_2=\sqrt{2\pi \pla}$ could be
neglected at energies 
\be
E\gg \frac{1}{\sqrt{2\pi \pla}},
\ee  
so we must worry about critical temperatures lower than $\beta_2^{-1}$ only. 
The first winding critical temperatures corresponds to the first excited
winding mode and its influence on the thermal behavior is suppressed at energies
\be
E\gg \frac{
\sqrt{\pla}R_{DD}
}
{\sqrt{8\pi \pla}
\lp \sqrt{\pla} R_{DD} -\lp \pla R_{DD}^2-1 \rp^{1/2} \rp }=E_H.
\ee 
Finally we will also consider the energies at which the winding states
become to appear and consequently the density of states that neglects it becomes
to fail. It occurs when $E > R_{DD}/\pla=E_{\omega} $. Along all these consideration we
have assumed the radii of all the Dirichlet directions to have the same
value $R_{DD}$.

We are now able to consider the phase space defined by
$R_{DD}$ and $E$ and to study the different regions that emerge from the 
previous conditions. These regions will be separated by the equality
relationship between $E_{\omega}$, $E_H$ and the energy of the system $E$. The
resulting phase space is sketched in Fig.(2). 
 
\begin{figure}
%
%
\let\picnaturalsize=N
\def\picsize{2in}
\def\picfilename{phase.ps}
\ifx\nopictures Y\else{\ifx\epsfloaded Y\else\input epsf \fi
\let\epsfloaded=Y    
\centerline{\ifx\picnaturalsize N\epsfxsize \picsize\fi
\epsfbox{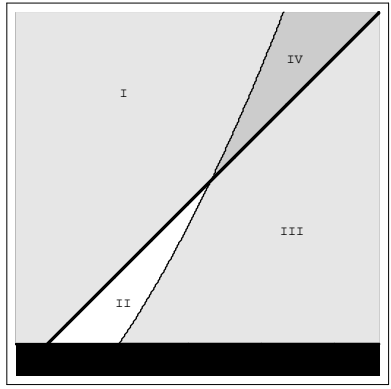}}}\fi
\label{phase}
\caption{{\small The phase space of the single string system that comes from the
Hagedorn dominance rule (\ref{hdr}) and the winding excitation energy
$E_{\omega}$}}
\end{figure}
 
 Suppose the system we are interested has
$DD-a$ Dirichlet dimensions compactified at radius $R_{DD}$, while the purely
Neumann and $a$ Dirichlet ones are already open.

If $E_H\ll E_{\omega}\ll E$ the winding states are excited when the 
Hagedorn term of the density of states clearly dominates the dynamics of
the system. As a consequence the winding influence could be
neglected. In this case the thermal behavior of the single string is
accounted by the first term of (\ref{omegaD}) with $\tilde{D}=NN+DD-a$. This
is in fact the situation in which the radius of the $DD$ direction is small
and could be better treated by tacking advantage of $T$-duality converting
the $DD$ directions into $NN$  ones. In the figure this regime correspond
the region $I$. 

In the case in which $E_H\ll E\ll E_{\omega}$, the region $II$ of the
picture, we have a system that
effectively lives in a space without the $DD$ directions and whose
thermodynamics is governed by the Hagedorn term. The density of states in
this case is the leading term of (\ref{omegaHighopen}). However the existence 
of this region is closely related to the value of the cut-off $\Lambda^2$, and
appears at energies $E\sim E_c$, so we think that it must be studied by the
IR regular  density of states.

We can go now to the region $III$ in which $E \ll E_H \ll E_{\omega}$. This
region behaves exactly in the same way as the region $II$, but the difference
comes from the fact that it appears for every value of the cut-off
$\Lambda^2$. On the other hand it must be described by the high energy
density of states in an $a$-dimensional background.

Finally we must look to the region $IV$. It corresponds to $E_{\omega}\ll
E\ll E_H$. In this case all the terms of (\ref{omegaD}) are relevant. We
think that it would be better accounted with a complete density of states,
overall when going to the multiple string system. The advantage of the
asymptotic expression over the exact one, that is simplicity, is lost and it
is better to use the complete series.

We would like to make two final comments on the thermodynamics at high 
energies. It is interesting to note that if we pass from the region $III$ to
the $I$ of the phase space, by increasing the energy of the system, it could test 
large extra dimensions. Then by this mechanism we can change the
dimensionality if the background in which the system is evolving.

On the other hand it seem remarkable to us that there are some
target background in which by this transition, 
we pass from a system which has a negative specific heat to another one  with
a positive and finite specific heat and which asymptotically ends at the Hagedorn
temperatures. In terms the nomenclature of \cite{barbo} we go from a 
non-limiting case to a limiting one.

Finally we would like to mention how these densities of states could apply
for the multiple string gas. Using the convolution description of \cite{us1}
we know that the multiple string gas density of states is obtained as
\be
\Omega(E)=\sum_{N=0}^{\infty}
\frac{1}{N!}\,\Omega^{(1)}(E)\star\cdots\star\Omega^{(1)}(E)
\ee 
where the stars stand for convolution integrals. One of the possible
approximations at high energy that we can do for the density of states of
the $N$-string gas is to consider it as behaving as a single long string. If
we use this picture we arrive at	
\be
\frac{1}{N!}\,\Omega^{(1)}(E)\star\cdots\star\Omega^{(1)}(E)\rightarrow
\frac{1}{N!} \lp A \frac{V^{NN}}{V^{DD}}\rp^{N}\Omega^{(1)}(E)
\ee
where $A$ is a factor that must be computed for each case. The simplest one
are those in which the single string is Hagedorn dominated, and we finally
obtain an ultra-high energy density of states as
\be
\Omega(E)\simeq
\frac{
e^{
\beta_H E+A 
\frac{V^{NN}}
{V^{DD}}
}}
{E^{(9-\tilde{D}-\nu)/2}} {\cal H} \lp E- E_c\rp. 
\ee
It corresponds to the leading order in the computation of \cite{deo,barbo}
and the corrections come from the series of $\Omega^{(1)}(E)$. The thermal
behavior of the multiple string system exactly matches with the single
string one with the previous assumption, as for $\Omega_1(E)$ it does not 
work in the IR region. We finally conclude that  only a complete computation 
of the single string density of states can be trusted for the multiple object 
gas.

\section{The Complete Density of States}

 In what follows we will obtain the complete density of state for the single
string and use it to describe the string gas thermodynamics. The idea here is to try to
inverse Laplace transform the whole free energy of the open string gas. 
The method that we will follow was presented in \cite{us1}. We write
the free energy of the open string theory as
\be
-\beta F(\beta)=-(2 \pla)^{-(NN+DD+1)/2} \frac{V^{NN}}{V^{DD}}\frac{\beta}{4
\pi}\sum_{\{M\}}\int_0^{\infty}\frac{dt}{t}t^{-(NN+DD+1)/2}e^{-\frac{\pi}{2t}
M^2} 
f(t).
\label{complete2}
\ee
where the internal partition function is taken as in (\ref{internal}). The
parameter $M^2$ is the mass of the closed string boundary states propagating
between the branes with the Euclidean time compactified on a circle of
length $\beta$.  It is given by
\be
M^2=\frac{1}{\pla} \lc \sum_{i}^{NN} n_i^2 R_i^2+ \sum_{i}^{DD}
\frac{m_i^2}{R_i^2}+\frac{\beta^2}{4 \pi^2} (n+1/2)^2\rc. 
\ee
At this moment we need to express the internal partition function as a sum
over open string states. More concretely we want to explicitly write
\be
f(t)=\sum_{k=0}^{\infty} a_k \,e^{-m^2_kt}
\hspace{5mm}
 \mbox{with} 
\hspace{5mm} a_k=\frac{1}{2\pi i}\oint
\frac{dq}{q^{k+1}}f(q).
\ee
Including the previous Taylor expansion and the parameter $\nu$ in
(\ref{complete2}), we can finally
write
\be
-\beta
F(\beta)=-(2 \pla)^{-(NN+DD+1)/2} \frac{V^{NN}}{V^{DD}}\frac{\beta}{4
\pi}\sum_{\{M\}}\sum_{k=0}^{\infty}a_k \int_0^{\infty}\frac{dt}{t}t^{-(10-\nu)/2}e^{-\frac{\pi}{2t}
M^2} e^{-k\pi t}.
\ee 
By using the integral representation of the Bessel functions $K(z)$  we can 
express the Helmholtz Free Energy of the open string gas as a sum over the Free
Energies of each target massive fields in the following way
\be
-\beta                                                        
F(\beta)\sim - \frac{V^{NN}}{V^{DD}}\frac{\beta}{2   
\pi}\sum_{\{M\}}\sum_{k=0}^{\infty}a_k 
(k\pi)^{\frac{10-\nu}{2}} 
K_{\frac{10-\nu}{2}} \lp \pi M
\sqrt{2 k} \rp 
\lp \pi M \sqrt{\frac{k}{2}}\rp^{\frac{\nu-10}{2}},
\label{bessel}
\ee
where the constant in front of the integral (\ref{complete2}) have been 
ignored.

At this moment we would like to obtain the single string density of states
by the inverse Laplace transformation of the Free Energy in (\ref{bessel}) as 
it was done in \cite{us1}. The complicate dependence of the argument of the Bessel function
on the compactification parameters, the radii $R$, and on the inverse
temperature $\beta$, makes this computation hard. We will attempt it by a
simple trick, that however seems useful to study some interesting and
simpler limit before. 

Let us begin with the study of the decompactification limit that corresponds
to taking all the radii to infinity. In this background the relevant degrees of
freedom would correspond to the momenta in the $NN$ directions. As we
discussed in section 3 this is the
only way to completely avoid the winding contribution. It is due to the fact
that we can forget  about the winding states for the single string by taking
its energy smaller than the squared Dirichlet radius
$R_{DD}$, but in order to do the same for the multiple string gas
we must prevent the statistical fluctuations that could allow a winding
nucleation. The density of energy is finally constrained to be in a low
range of values by (\ref{nowindingenergy}).  In summary, the correct way to study the
system in absence of winding states consists of taking the radius of the $DD$ 
direction to infinity and not to assume string energies lower than the winding
masses.

We take the desired limit in (\ref{complete}), by assuming the $NN$ momenta
to have a continuous spectrum and taking only into account the winding zero
mode we arrive at
\be
-\beta                                                        
F(\beta)\sim - V^{NN}\frac{\beta}{2   
\pi}\sum_{n\in \z}\sum_{k=0}^{\infty}a_k 
(k\pi)^{\frac{NN+1}{2}} 
K_{\frac{NN+1}{2}} \lp \pi M_n
\sqrt{2 k} \rp 
\lp \pi M_n \sqrt{\frac{k}{2}}\rp^{-\frac{NN+1}{2}},
\label{besselopen}
\ee
where the masses now are reduced to
\be
M_n=\frac{\beta |2n+1|}{4 \sqrt{\pla} \pi}.
\ee
The density of states corresponding to this free energy can be obtained by
using exactly the same method of \cite{us1}, getting the following result
\be
\Omega^{(1)}(E)=\frac{2 V^{NN}
(4\pi)^{\frac{NN}{2}}}{\Gamma(\frac{NN}{2})}\sum_{n=odd}\sum_{k=0}^{\infty} a_k  
 n^{-(NN+1)} E \lp E^2-\frac{n^2 k}{8\pla} \rp^{\frac{NN}{2}-1}{\cal H}\lp
E^2-\frac{n^2 k}{8\pla} \rp. 
\label{nowinding}
\ee

We can compare the high energy density of states for the single  string
system in the open universe case, that corresponds to the result expressed
in (\ref{omegaHighopen}). The crucial difference comes from the IR behavior. 
The expression in  (\ref{nowinding}) is regular at low energies  because we did not exclude out any mass
levels of the string, as we did for the first case, by using $\Lambda^2$. 
The critical Hagedorn phenomenon  here is hidden in the growth of the
coefficient $a_k$, and appears as the energy is big enough to open  massive
string  thresholds. 

We can compute the density of states for the string gas living in a universe
with compact dimensions. The way of computing it is the same we have used
above. In this case the free energy is represented by the $K_{1/2}(z)$
Bessel function. The trick that we will use here to use the result of
\cite{us1} and obtain the density of states consists of writing the free
energy in (\ref{complete2}) as 
\be
-\beta F(\beta)=-(2 \pla)^{-(NN+DD+1)/2}
\frac{\beta}{4\pi}
\sum_{n=0}^{\infty}
\sum_{\{ M_{knw} \} }
\int_0^{\infty}
\frac{dt}{t}
t^{-1/2}
e^{-\frac{\beta^2}{8 \pi t} (n+1/2)^2}
e^{-\frac{2\pi}{\pla}M_{knw}^2} 
\ee
where the mass threshold here are expressed in terms from the open string
side as 
\be
M_{knw}^2=\sum_{i=1}^{NN}\frac{4 n_i^2}{R_i^2}+\sum_{j=1}^{DD}4
w_j^2 R_j^2+2\frac{\pi}{\pla}^2 k.
\ee

The case in which all the dimensions are compact and their radii are small enough 
to force us to consider the spectrum of momentum and winding  
strongly discretized, is very subtle. We will then begin with the case in which 
we  assume that the $NN$ direction are open and the $DD$ ones are 
compactified before. The density of states in this case is 

\be
\Omega_{(1)}(E)=\frac{2 V^{NN}
(4\pi)^{\frac{NN}{2}}}{\Gamma(\frac{NN}{2})}\sum_{n=odd}\sum_{M_{kw}}^{\infty} a_k  
 n^{-(NN+1)} E \lp E^2-\frac{n^2 M_{kw}}{16} \rp^{\frac{NN}{2}-1}{\cal H}\lp
E^2-\frac{n^2 M_{kw}}{16} \rp. 
\label{windingonly}
\ee
where the string masses now are related to the winding number only  
\be
M_{kw}^2=\sum_{j=1}^{DD}4
w_j^2 R_j^2+2\frac{\pi}{\pla}^2 k.
\ee
With this density of state is very easy to study the case in which the
single string cannot probe some $DD$ direction. It is simply accounted by
taking some of the winding radii to infinity, another important feature is
that the IR validity of it allows us to study the multiple string gas and
then we would be able to analyze the possibility of defining some effective
dimensions and the existence of energy density scales that probe extra large
compactified directions.

Finally we are able to obtain the density of states, for the a null number
of non-compact $NN$ dimensions. It corresponds to 
\be
\Omega^{(1)}(E)=\frac{2}{\Gamma(\frac{0}{2})}\sum_{n=odd}
\sum_{M_{knw}}^{\infty} a_k  
 n^{-1} {\cal \delta}\lp
E-\frac{n\, M_{knw}^{1/2}}{4} \rp 
\label{both}
\ee

The first noticeable feature of this result is that we do not have any
continuous dependence on the energy. The density of states above strictly
corresponds to the degeneracy of each mass level without any kinematical
contribution. On the other hand we know that by $T$-dualizing the $NN$
directions we can describe the system of wrapped D-branes as a system of
D-particles on a torus. It is interesting to note that if we take the radii
of the $DD$ direction to infinity we arrive at a system of D0-branes in an
already open target space. In this case the density of states is reduced to
a sum of the mass degeneracy times the Dirac-delta function located at each
mass level. 

Two relevant limits of this result  should  be analyzed. If we take the
energy to be large one we should arrive at a result related to the 
exponential growth of the degeneracy of string mass levels. Doing it in
(\ref{both}) we in fact arrive at 
\be
\Omega^{(1)}(E=m)\sim e^{\beta_H m}\,m^{-9/2}{\cal H}\lp m-\frac{2\pi}{\beta_H
\Lambda^2} \rp,
\ee
that corresponds to the leading term of (\ref{omegaHighopen}) and to the
critical behavior studied in the early works on the subject  \cite{hag,carli}.
As we mention before, because of the absence of any kinematical degrees of
freedom, the energy here is exactly the mass of the strings.
What  is remarkable is that once again the cut-off establish the range of
validity of the approximation. 

On the other hand it is interesting to relate the previous result to the SYM
description of the D-brane thermodynamics \cite{green,vaz1,tse} and more
concretely to the finite temperature behavior of Matrix model of M-theory
\cite{BFFS,therm}. Taking the massless limit of the open string theory on the
brane we arrive to the SYM description of the system. On the microcanonical
side this is implemented by simply taking into account the first term of the
density of states above. It finally gives a constant density of states that
equals the polarization of the massless string modes times a delta at zero
energy. We can obtain the same
result by firstly starting by the canonical free energy of the SYM$_{0+1}$
theory. It reads 
\be
F(\beta)\sim \beta^{-1}\sum_{n=odd}^{\infty}n^{-1}.
\ee
If we make the inverse Laplace transformation of 
$Z^1(\beta)=-\beta F(\beta)$, we easily see that the single-object density
of states $\Omega^{(1)}_{SYM}(E)$ exactly
matches the massless limit of our result in (\ref{both}).

One final comment. The  density of states in (\ref{both}) has the divergent
$\Gamma(\frac{0}{2})$ in the prefactor. The value of this function makes the
density of states vanish and this needs some interpretation. What is
happening is that, because of the compactness of the background universe,
the spectrum of the string mass is discrete.  This means that in a given
energy range there could be isolated accessible states only, in such a way
that we cannot define a density of states. In our computation we have
supposed the spectrum to be continuous and we recover a vanishing
$\Omega^{(1)}$ for the single string. In principle thermodynamics is not
affected  because, in the microcanonical ensemble, the physical magnitudes
come from the derivatives of the entropy, which is proportional to the
logarithm of the density of states, so we can anyway neglect any
multiplicative constant.

\section{The number of strings and the entropy.}

   In \cite{us1,us2,tan,deo,lee} a picture was studied for the phase
transition occurring at the Hagedorn temperature. The transition signal was
the appearance of a \cal{fat string} that breaks equipartition and absorbs
enough energy to excite high mass levels while all the others remain
massless and form a kind of background, that we called 'the sea'
\cite{us1,us2}. We would like now to see how the number of strings evolves
and how the sea and the fat string share the energy as the volume varies. To
study that, we approximate the high-energy density of states by

\be
\Omega_{High}(E,E_{sea},V)=V (E-E_{sea})^{-\frac{a}{2}} e^{\beta_H (E-E_{sea})}
\Omega_{sea}(E_{sea})
\label{sea}
\ee

 with
\be
\Omega_{sea}=\frac{1}{\Gamma(d N)} \frac{1}{N!} \lp \frac{\Gamma(d)
\zeta(d+1)(d-1)^2}{2^{d-1} \pi^{d/2}} f(V) \rp^N E_{sea}^{d N-1}
\ee

 where $a$ is a constant integer that depends on the model and the D-brane
background, $d$ is the number of open dimensions (testable and continuous at
low energies), $N$ is the number of strings and $E_{sea}$ is the mean energy
of the sea. We can use concepts like average energy in spite of working with
the microcanonical ensemble because we have separated our microcanonical
system into two parts. Each possesses a thermodynamical (physically
infinite) number of degrees of freedom so that they act as an 'Energy
reservoir' for the other. That is, the sea of massless strings can be
thought of as a gas in the canonical ensemble. Throughout the following
calculation we shall ignore the thermal fluctuations, so that we shall
suppose that the energy or the volume are large enough.

   Firstly, we calculate how the energy is shared between both subsystems.
The most probable distribution is given by the following equation:
\be
\left. \frac{d \Omega_{High}}{d E_{sea}} \right|_{E,V}=0
\ee

   The solution to it is
\be
E_{sea}=\frac{d N}{\beta_H},
\ee
each light string having an average energy $d/\beta_H$, which is the energy
at which the massless fields reach the Hagedorn temperature in $d$ spatial
dimensions. This result was already obtained in \cite{us1} using a model
of kinetic theory.

   This is part of the equation of state and it allows us to write
\be
\Omega_{High}(E,E_{sea},V)=\Omega_{High}(E,N,V).
\ee

   This form of the density of states permits to calculate the most
probable average number of strings for given values of the total energy and
the volume and therefore, how the energy is shared between the two
subsystems. The equation is:
\ber
\left. \frac{d \Omega_{High}}{d N} \right|_{E,V} & = & 0 \nonumber \\
E_{fat}= \lp E-\frac{d N}{\beta_H} \rp & = & \frac{d a}{2 \beta_H}
\lc -1 - log \lp \frac{\Gamma(d) \zeta(d+1) (d-1)^2}{2^{d-1} \pi^{d/2} N }
f(V) \rp \rc^{-1}
\label{crit}
\eer

   For certain values of $V$ and $N$ the energy of the fat string
becomes small or even negative, that is, it ceases to exist. This gives a
critical
value for $f(V)$ that separates two phases with and without the fat string.
The phase without fat string is easy to analyze because the main
contribution comes from the massless string gas. Its density of states is
\be
\Omega(E,V)=\sum_{N=0}^\infty \frac{1}{\Gamma(d N) N!} \lp \frac{\zeta(d+1)
\Gamma(d)(d-1)^2}{2^{d-1} \pi^{d/2}}\rp^N V^N E^{d N-1}
\ee

   For large volumes and energies, it yields
\ber
T=\frac{1}{d}\epsilon \nonumber \\
\epsilon= d \lp \frac{2^{d-1} \pi^{d/2} }{\zeta(d+1) \Gamma(d) (d-1)^2}
\rp^\frac{1}{d} \rho^\frac{1}{d}
\label{eos}
\eer
where
\ber
\epsilon=\frac{E}{<N>} \nonumber \\
\rho=\frac{<N>}{V}
\eer

  The subsequent analysis of this expressions depends on the particular
form of $f(V)$. In the case of closed strings it is $V+1/V$, that tends to
$V$ away from the selfdual volume. Besides, $d$ equals $9$ and $a$ equals
$1/2$. As a result, equation (\ref{crit}) gives a critical value of
$\rho_{crit}\simeq 159$ strings per $\alpha'^{9/2}$. If we now take the
thermodynamical limit ($V$ and $E$ $\rightarrow\infty$) $N$ dynamically
diverges maintaining the densities finite. As both densities, $\rho$ and
$\epsilon$, grow together, any of them can be taken as the only parameter
defining the system. The physical picture is as follows: If the density is
lower than the critical one, the system behaves like a massless gas in ten
dimensions and the temperature grows linearly with $\epsilon$. When the
density reaches the critical value a phase transition occurs and, as the
correlation length tends to infinity, the system tends to concentrate the
energy on one fat string with infinite mass. Beyond the transition point,
the energy of the fat string becomes finite and decreasing with the density.
The finiteness of $E_{fat}$ drive us to formally say that its 'Energy
density' is zero. We use the inverted commas because one can hardly talk
about a density for a single object but this is quite a peculiar one because
it has influence over global properties. Its r\^ole is to regulate the
temperature so that the bath always remains at $T_H$. The system prefers to
create more and more light strings rather than exciting higher massive modes
of the already existing ones.

   The evolution of the temperature is singular. At low energies, the
temperature grows with increasing specific heat. At the critical point, the
temperature is $T_H$ and remains so thereafter with infinite specific heat.
This is a confirmation of the image depicted in \cite{us1}.
Fig.(5) shows the qualitative behavior. 

 The case of open strings attached to D-branes is quite different because
\be
f(V)=\frac{V_{NN}}{V_{DD}}.
\ee
The transition depends on how we take to infinite both volumes. If
\be
\frac{V_{NN}}{V_{DD}}\rightarrow \infty
\ee
the analysis is identical to that of closed strings, but defining a new
density
\be
\rho=\frac{N V_{DD}}{V_{NN}}
\ee
   It is the density of strings one would define in the T-dual theory where
all the directions have Neumann boundary conditions. Other possible limit is
the opposite one
\be
\frac{V_{NN}}{V_{DD}}\rightarrow 0
\ee
which is equivalent to taking $\rho$ to infinite. The energy of the fat
string tends to zero but is still acts as a regulator so that there is an
equilibrium between an unstable fat string and the sea that fills the space
at $T_H$.

  There is one exceptional case when $a=0$, that is, for open strings in a
completely testable universe. In that case, the breaking of
equipartition is quite peculiar. Compare figs. (3) and
(4). There is never a sharp peak in the density of states of
several strings signaling the most probable appearance of a fat one that
absorbs much more energy than the others. Rather, there is a spread in the
probability that means that the fluctuations in the energy of a single
component are comparable to the total energy. This invalidates the physical
image of the single string dominance of thermodynamics and even prevents the
correction of it due to what we have called the sea. The correct physical
image is a gas where equipartition more or less holds but with very large
fluctuations in the energy of particular strings. This makes the need for
the exact calculation stronger because the probability of finding a string
in virtually any point of the phase space is large. The only prediction that
can be obtained with the asymptotic expressions is that the asymptotic
temperature is Hagedorn and that the specific heat is always positive.

   This is the only case in which there is not a phase transition if we take
the appearance of the fat string as the order parameter. However, at energy
densities beyond the Planck scale, the system acquires a rather anomalous
behavior. For the moment, we cannot precise if it corresponds to a higher
order phase transition or simply a soft change of behavior.

   The difference between the usual microcanonical point of view and this
one is the fact that here we are letting the number of strings dynamically
vary, that is, we are working in the ($E$, $V$, $\mu$)-ensemble instead of
the pure microcanonical. The former is more adapted to the string gases
because they naturally have a fixed and null chemical potential. Somehow, it
is a way to gather the different descriptions in terms of finite-$N$ gases
together. As explained in sections $4$ and $5$, one can make convolutions of
the single string density of states and obtain the behavior of different
subsystems with a fixed number of constituents. The ($E$, $V$,
$\mu$)-ensemble shows us how the transition between them happen.
 
\begin{figure}
%
%
\let\picnaturalsize=N
\def\picsize{2in}
\def\picfilename{equi2.ps}
\ifx\nopictures Y\else{\ifx\epsfloaded Y\else\input epsf \fi
\let\epsfloaded=Y    
\centerline{\ifx\picnaturalsize N\epsfxsize \picsize\fi
\epsfbox{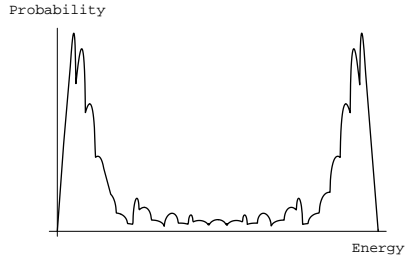}}}\fi
\label{diente}
\caption{{\small Probability of the energy distribution for 2-string system
in the $\nu=6$ case, at high energies.}}
\end{figure}

   The inclusion of the sea as an important contribution to thermodynamics
has consequences on the estimation of the entropy. One has to add the
logarithm of (\ref{sea}), that is, in the large-$N$ limit
\be
S(V,E,N)=log \Omega(V,E,N)=10 N+ N log \frac{C V E^9}{9^9 N^{10}}
\ee
where $C$ is a constant. Substituting the value of $N$ that solves the
equation of state (\ref{eos}), it is
\be
S(V,E)=10 N= \frac{10}{9} \rho V=\frac{10}{9} \lp \frac{\epsilon}{9} \rp^9 V
\ee

  This is the leading contribution except when $a=0$.

  Let us now write about the pressure. The density of states that we have
obtained in the previous section has the functional form:
\be
\Omega(V,E)=f(E) \times g(V)
\ee
where
\be
g(V)=\frac{V_{NN}}{V_{DD}}
\ee

  This means that the dependence of the pressure on the temperature is the
same as that of the free energy. However, its dependence on the volume is
not trivial at all. Let us recall that, in traditional theories
\be
P=\left. -\frac{\partial F}{\partial V} \right|_T.
\ee

   However, we know that T-duality is an exact symmetry in type II String
Theory. This means \cite{us2} that every physical (measurable) magnitude
must be invariant under T-duality. This, of course, includes the pressure.
The usual definition, that we have written just above, is not T-selfdual and
therefore it must be corrected and substituted with another. We shall use
the proposal in \cite{us2}.

\be
P=\left. -\frac{\partial F}{\partial V} \right|_T {\cal H}\lp V-
\ls^{\tilde{D}/2}
\rp -\left. \frac{\partial F}{\partial (1/V)} \right|_T {\cal H}\lp
\frac{1}{V}- \ls^{\tilde{D}/2}  \rp
\ee

   This case is even more complicated because the function we are interested
in is not homogeneous in the volume, so the derivative must be decomposed as
 
\ber
V=\prod_{i=1}^{\tilde{D}} (2 \pi R_i) \nonumber \\
dV=\sum_{i=1}^{\tilde{D}} \frac{V}{R_i} d R_i \\
\frac{\partial}{\partial V}=\sum_{i=1}^{\tilde{D}} \frac{R_i}{V}
\frac{\partial}{\partial R_i} \nonumber
\eer
and
\be
P=\sum_{i=1}^{\tilde{D}} \lc \left. -\frac{R_i}{V} 
\frac{\partial F}{\partial R_i} \right|_T {\cal H}
\lp R_i- \ls^{1/2} \rp \left. -\frac{1}{V R_i}
\frac{\partial F}{\partial 1/R_i}
\right|_T {\cal H} \lp \frac{1}{R_i}- \ls^{1/2} \rp \rc
\ee
with this, and if we assume that all $R_i$ are larger than $\ls$,
\be
\frac{d}{d V}\frac{V_{NN}}{V_{DD}}= \frac{1}{V_{DD}^2} (NN-DD).
\ee

   It is interesting to see what happens when we make T-duality act on the
free energy so that, for example, all the directions are Neumann, the result
would be
\be
\frac{d}{d V} V_{NN}^{\mbox{\tiny T-dual}} = (NN-DD),
\ee
where $NN$ and $DD$ are the original Neumann and Dirichlet dimensions. The
result is similar up to a multiplicative constant term. In fact one can see
that what really coincides in both points of view of the same theory is the
product
\be
P V \propto (NN-DD) \frac{V_{NN}}{V_{DD}}
\ee

    One remarkable characteristic of the pressure we have calculated is
that, in certain cases, it vanishes. This happens whenever the number of
directions with Neumann and Dirichlet boundary conditions coincide. This is
only possible if an odd number of dimensions is untestable.
 
\begin{figure}
%
%
\let\picnaturalsize=N
\def\picsize{2in}
\def\picfilename{equi.ps}
\ifx\nopictures Y\else{\ifx\epsfloaded Y\else\input epsf \fi
\let\epsfloaded=Y    
\centerline{\ifx\picnaturalsize N\epsfxsize \picsize\fi
\epsfbox{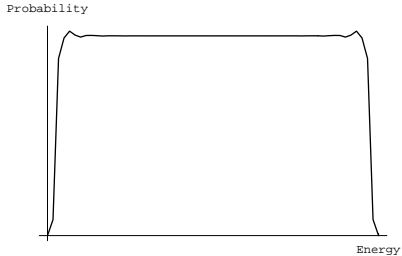}}}\fi
\label{equi}
\caption{{\small Probability of the energy distribution for 2-string system
in the $\nu=0$ case, at high energies.}}
\end{figure}

\section{Conclusions}

   In this paper we  have studied the thermal behavior of string systems in 
D-brane backgrounds in the microcanonical ensemble. The goal of this kind of
analysis is to delucidate what is the r\^ole of the Hagedorn temperature in
the thermodynamics of Dirichlet open string theories. As in the closed
string cases the necessity of the microcanonical description comes from the
breaking of the canonical ensemble at Hagedorn temperatures.  

Our strategy has been as follows, and is along the lines of \cite{us1,us2}. 
We have begun by a computation of the asymptotic density of states of the
single string. These computation are analogous to those of \cite{tan,deo} for
the closed-string gas and have been recently used for the D-brane cases in
\cite{barbo}. We have shown that this procedure  corresponds to including an IR
cut-off $\Lambda^2$ that eliminates the contribution of the massless string
modes. Consequently the densities of states obtained are not valid for the
low energetic regime.  We have studied the range of validity of this
approximation in terms of the energy of the system, the radii of the compact
dimensions and the value of $\Lambda^2$.  The result is a kind of phase
space where different parts are best described by particular approximations.
Only two of these parts are well described by the asymptotic density of
states, approximated by the Hagedorn term only. Passing from one of them to
the other the number of effective dimensions of the target space the
string gas lives on changes. We claimed that in order to completely know the thermal
behavior it is necessary to use a complete, IR regular density of states.

On the other hand the thermodynamics of the string gas is not exactly
described by the single string system, so we need to study the multiple
string density of states. We should follow the convolution prescription of
\cite{us1}. However it is not possible to do it using the high energy
densities of states of the single string because of their irregular IR
behavior. Finally one is lead to look for a complete density of states.

Before computing the desired density of states we have analyzed the thermal
properties of the single string. We have shown that there is a negative
specific heat phase in all cases except for the $\tilde{D}+\nu=9$ one. In the latter
we have an IR definite phase but we think that at these energies the density
of states does not include all the information. It is in fact obvious
that the corrections to the leading behavior are strongly dependent of
$\Lambda^2$,and that way nothing warrants that we are including all the IR
degrees of freedom of the theory. The only extractable conclusion is that
after including all the IR information we will still have a positive
specific heat phase, and Hagedorn will be a asymptotic maximum temperature
of the system. 
 
We have calculated an exact expression for the single string density of
states in section 5. It is got in the form of a series with a finite number
of terms for finite values of the energy. The results are analogous of those
in \cite{us1,us2}. With those instruments, we are able to study the
thermodynamics of the systems and also to match our results with those
obtained using the SYM$_{p+1}$ description of the D-branes dynamics. We have
used the $\tilde{D}+\nu=9$ cases as some special examples.  In one of
them, $DD=9$, we are able to connect with the Matrix model microcanonical
description, in the Born-Oppenheimer approximation.

In particular, we study the distribution of energy among the constituents.
This gives us physical images that help us test models that had been
proposed in earlier works \cite{us1,us2,tan}. When a fat string with more
energy than the rest appears and, therefore, equipartition is broken, a
phase transition occurs. We have been able to distinguish the cases in which
this happens and in which it does not. We have seen how this affects the
temperature and the specific heat value, the case in which the phase transition
occurs has been plotted in Fig.(5). When several of this systems are put in
thermal contact, that is, when open strings attached to different D-branes
and free closed strings thermally interact, the behavior of the global
system is dominated by the most degenerate phase. This is the one that gives
lower temperature and smaller specific heat for a value of the energy: open
strings with $\tilde{D}+\nu=9$. A similar picture has already been obtained in
\cite{barbo}; the difference between our picture and theirs is that in ours
all subsystems have $T_H$ as the maximum temperature. There is not any
'non-limiting' system with temperatures higher that $T_H$. Our result is
that systems can be classified between those with phase transition and those
without it. The systems without phase transition are the closed strings in a
completely compactified space \cite{us2} and open strings with
$\tilde{D}+\nu=9$. Whenever there is at least one open direction, the only
system that does not suffer phase transition is that of open strings with
$\tilde{D}+\nu=9$. When one considers the whole system made up with closed
strings and open strings of various types (various values of $\nu$,$DD$ and
$NN$), the most degenerate system dominates the thermodynamics. It is always
 the one without transition. This means that the existence of one D-brane
prevents any phase transition, except when some $DD$ are open.

Along all these considerations we have considered all the untestable $DD$
direction to be already open. If we had not done it we would have found some
of the previous systems in thermal contact having the phase transition. It
corresponds to systems that stand in region $III$ of fig.(2). This systems
have the energy per string ($\rho_N$) constraint by (\ref{nowindingenergy}).
On the other hand we know that the phase transition occurs at $E/N \sim
\tilde{D}\,T_H=\rho_H$. Two possibilities come. If $\rho_H<\rho_{\omega}$
the systems suffers a phase transition. Else we need to include all winding
states and finally we pass to the region $I$ where we do not find any phase
trasition. We think that in the thermodynamic limit, where $\rho_{\omega}$
goes to zero, a phase transition is prevented.
\begin{figure}
%
%
\let\picnaturalsize=N
\def\picsize{2in}
\def\picfilename{tentati.ps}
\ifx\nopictures Y\else{\ifx\epsfloaded Y\else\input epsf \fi
\let\epsfloaded=Y    
\centerline{\ifx\picnaturalsize N\epsfxsize \picsize\fi
\epsfbox{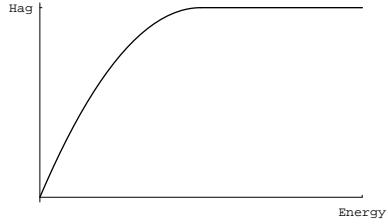}}}\fi
\label{tentati}
\caption{{\small Temperature versus energy for the multiple string systems
that reach Hagedorn temperature through the long string nucleation}}
\end{figure}

Regarding the pressure, we calculate it in several cases and see the
peculiarities of string systems using the String-Theory definition of it
proposed in \cite{us2}.

  Finally, we use the 'fat string-sea' model to make a calculation in a
genuinely thermodynamic limit, and see the evolution of the temperature as a
function of string and energy densities instead of absolute magnitudes.

\section{Acknowledgments}

   We thank a lot M. A. R. Osorio  for enlightening and very helpful
discussions and J. L. F. Barb\'on for useful comments. The work of M. L. M. 
was partially supported by MEC under a
FP-97 grant.

\end{document}